# On Critical State Transitions between different levels in Neural Systems

Gerhard Werner
Department of Biomedical Engineering
University of Texas , Austin TX, 79731, USA.
gwer1@mail.utexas.edu

The framework of "Modern Theory of Critical State Transitions" (1, 2) considers the relation between different levels of organization in complex systems in terms of Critical State Transitions. A State Transition between levels entails changes of scale of observables and, concurrently, new formats of description at reduced dimensionality. It is here suggested that this principle can be applied to the hierarchic structure of the Nervous system, whereby the relations between different levels of its functional organization can be viewed as successions of State Transitions: upon State Transition, the 'lower' level presents to the 'higher' level' an abstraction of itself, at reduced dimensionality and at a coarser scale. The re-scaling in the State Transitions is associated with new objects of description, displays new properties and obeys new laws, commensurate to the new scale. To illustrate this process, some aspects of the neural events thought to be associated with Cognition and Consciousness are discussed. However, the intent is here also more general in that State Transitions between all levels of organization are proposed as the mechanisms by which successively higher levels of organization "emerge" from lower levels.

**Introduction**

The functional architecture of neural systems extends over a range of up to five orders of magnitude of scales in space and time: from microns and microseconds for ion channels at one end, to centimeters and tenth of seconds for interareal neuron clusters, at the other; and a 100 fold range of frequencies in the wave domain. In actual practice and on methodological grounds, this range is generally addressed at three levels of inquiry: microscopic (neuron spike activity), mesoscopic (local field potentials) and macroscopic (electroencephalogram, brain imaging) activity (3,4). Conceptually, the functional relations between and within levels of organization are often considered in terms of integration and differentiation (5,6) or with a more specific functional connotation as emergence versus downward causation, respectively. Alternative conceptualizations are formulated in terms of the larger scale being 'slaved' to the smaller by some coupling function or circular causality (7). The wide range of scales poses considerable conceptual and practical difficulties for designing large scale brain models, as does the scale-dependent change of degrees of freedom from level to level. Breakspaer and Stam (8) illustrate the latter relation in numerical models of between-scale bifurcations. Brain modeling across multiple scales was also the subject of inquiry by Robinson et al (9). Honey et al (10) explored issues of functional connectivity across multiple time scales. Horwitz and Glabus (11) addressed the complex issues raised by the multiple spatial-temporal scales of neuroscience data. Much as the importance of scale differences is appreciated, I suggest that its implications are not generally taken into account: that is the realization that the scale on each level of observation can emerge from the next finer level by ignoring some of the lower level details which become irrelevant at the higher level (12,13,14). In this sense, scales can 'create' classes of objects of observation, each class having its distinct description and obeys its own laws. This point of view is being fruitfully applied for modeling spatiotemporal dynamics of systems in Ecology (15). In the following, I will associate scale transitions with concepts and principles of Critical State Transitions in 'Modern Critical Theory' (1,2). These principles are intended to be applicable to all inter- level relations, but I will illustrate their application here as a demonstration of concept for the relation between the neural events of Cognition and Consciousness. As a paradigmatic inter-level relation, it offers at this point relatively transparent criteria and benchmarks for applying the theory of Critical State Transitions, and what implications accrue from it.

The essay is organized in the following manner: a brief outline of the conceptual framework of the 'Modern Critical Theory" precedes a sketch of neural events in Cognition and Consciousness, limited to those aspects that relevant for applying the notions of State Space. The final Section deals with the implications of viewing inter-level relations as State Space Transitions. This is intended to lend credence to the view that critical state transitions segregate reality into a hierarchy of domains, each characterized by its intrinsic scale, and associated structure and organizational laws.

### The theoretical framework: Modern Theory of Critical State Transitions.

The framework applied in this essay is rooted in some aspects of Contemporary Physics, specifically the field of Critical State Transitions, as viewed in "Modern Critical Theory" (1,2). It is concerned with states of matter as organizational phenomena: elementary physical laws have in principle the ability to generate states and state transitions as organizational phenomena. (Be aware of terminological pitfalls: here and in the following , I use the term "states" to designate what in some other sources are also designated "phases"). The class of state transitions under considerations in this essay are characterized by the abruptness of their occurrence at critical points of their system parameters. Typically, the operation of local microscopic laws results in the formation of aggregates by coarse graining. For an illustration of this process, see Ref. 17. The collective state of matter is then unambiguously characterized by behavior that is exact in the large aggregate. The important feature of the organization following state transition is the bringing into existence of new objects with distinct properties. Starting with familiar examples : think of the switch from ferro- to paramagnetism, or the transition from water to ice. Typically, one deals with a large collection of 'microscopic' constituents which at state transition to a 'macrostate' display qualitatively novel features and properties. The macrostate's novel properties have no referent at the microscopic level and require new descriptors: think of hardness or liquidity in the ice-water example as descriptors of new physical referents. Accordingly, state transitions create new physical states which call for new descriptions of physical reality (16). In addition, "Modern Critical Theory" considers reality as a hierarchy of levels, each level having its own scale, its own description, and a theory that accounts for the scale of that description. The scale on each level emerges from the scale on the next finer level by ignoring some of the lower-level details which become invisible at the scale of the higher level (13,14): loosely speaking, view the macrostate as an abstract rendition of the microstate which recedes to invisibility in the new scale, comparable to changing to glasses with lower magnification.

Most of the successful applications of this framework come from Thermodynamics and Statistical Mechanics at or near critical state transitions of physical matter. As an empirical fact, certain parameters of material systems assume at the point of state transition critical values which define entire classes of system (Universality Classes), presenting identical macrostates despite a wide range of differences at the level of microscopic constituents (in terms of composition, physical properties etc). A corollary of this is the multiple realizability of emergent new macroscopic phenomena despite microscopic diversity (for details, see: Ref. 18). This principle can be illustrated with a metaphor taken from Probability Theory: when sets of multiple independent population samples are subsumed under some statistical (say Gaussian) Distribution, then the parameters of the distribution characterize a (kind of) Universality Class , i.e. the ensemble of which the individual sets of samples are independent realizations.

Universality and multiple realization designate that 1) some details of the system which would figure in a detailed causal-mechanistic explanation of the system's behavior, are in the limit to some extent irrelevant for characterizing the macroscopic phenomenology of interest; and 2) different systems with vastly different "micro" details can exhibit identical behavior at the macroscopic level. The result in both cases is a drastic reduction of the dimensionality in the state transition. State space transitions in physical matter can be simulated in many different kinds of computational models which permit examining the underlying events in detail. Procedures of "coarse graining" (19) and the strategies of Renormalization Group Theory (20) are the principle tools; the latter essentially, re-scaling by successive coarse graining while maintaining self-similarity.

### On the neural events in Cognition and Consciousness: setting the stage for a demonstration of Concept.

Here I merely sketch in broad outline the neurophysiological events that form the benchmarks for illustrating the application of Critical State Transitions. Implicit is the Dynamical System approach in Neurodynamics (for a recent discussions, see Refs. 21, 22). Abundant evidence for the brain's propensity to undergo State Transitions was secured by Freeman and Holmes (23), and by Freeman in numerous studied, see for instance Refs. 24, 25,26, extending also into considerations of quantum field theory (27). Jirsa et al (7, 28) studied abrupt transitions between sensorimotor coordination patterns in a general framework of multistability and switching in biological systems, applying the conceptual tools of Synergetics (29). Convincing evidence for brain state transitions is also summarized in publications by Bressler and Kelso (30) and Chialvo (31), and demonstrated by Gervasoni et al (32). At the mesoscopic level, Fujisawa et al ( 33 ) demonstrated that internal states of neurons (expressed in UP-DOWN alternation) can be subject to state transitions of their embedding recurrent neuronal networks.

The neural processes of the reactive behavior of organism's sensory-motor interaction with the environment are for this demonstration of concept taken at the microscopic level those constituting the neural events thought to be associated with consciousness as the macroscopic level. I adopt Searle's (34) notion of the state of wakefulness as basal (background) consciousness, a kind of unified field, presumably identical or overlapping with the condition of vigilance in the terminology of Dehaene & Changeux (35). Specific sensory events would then punctuate, as it were, the steady state of the unified field, as the basis of discrete subjective experiences. This notion of discrete events in Consciousness tallies with observations of Fingelkurts & Fingelkurts (36) of discontinuities in the EEG which they identify as transient operational brain microstates, signaling shifting activation of neuronal networks; and of Lehmann and associates (37) who describe punctuated abrupt changes in EEG activity as evidence of distinct steps in mental information processing.

Comparing neural models with perceptual phenomena, Dehaene and Changeux (38) emphasize the suddenness of the transition to conscious and reportable registration of stimulus events, associated in the neural model with a "self-amplifying recurrent activity" in widely-distributed cortical regions. Edelman (39) considers dynamic reentrant interactions across cortical circuits as the medium for synchronous linking and binding among widely distributed brain areas. Sergent & Dehaene (40) and Del Cul et al. (41) take their findings with the attentional blink test and backward masking, respectively, to be concordant with the notion of a discrete threshold for access to consciousness. Abruptness of onset of conscious experience is also an essential aspect of the extended psychophysical studies of Breitmeyer and Ogman (42). Damasio (43) postulated synchronous activation of globally distributed convergence zones to serve as the neural substrate of recall and recognition, and subsequently. However, what appears to be also required for conscious experience is that the somatically embodied and environmentally reactive behavior be associated with the body's moment-to-moment adaptive bioregulatory processes, originating with a multiplicity of subcortical brain structures (44,45). In its totality, existing evidence supports the basic idea of Baars ' Global Workspace Theory' of associating consciousness with widespread access among otherwise independent brain functions (46,47). A particular view is advocated by a group of investigators who attribute the distinction between unconscious and conscious vision to recurrent processing: Lamme and associates (48,49) claim that reportable conscious visual experiences require that the "feedforward sweep" of neural activation from visual towards motor areas become extended to a "backward sweep" which consists of widespread recurrent activation of frontal, prefrontal and temporal cortex (50,51).

The foregoing sketch of observations and conjectures is intended to underscore the signal feature of neural events in Cognition and Consciousness that warrant considering a Critical State Transition: abruptness of change in large-scale patterns of connectivity of otherwise disparate brain regions. In what way the neural state transition may be associated with the emergence of mental states is outside the scope of this essay.

**Taking State Space seriously**

The context of the observations of the foregoing section and the totality of findings with Brain Imaging and Electroencephalography supports viewing the brain as a dynamical system of unprecedented complexity. This suggests adopting the State Space approach for characterizing the brain's states as points (or circumscribed regions) in a high dimensional space. The dimensionality of the state space reflects the number of the system's independent variables which can also be considered components of a state vector. A system's State Space encompasses the set of all potentially accessible states of the system: Dynamical system theory is concerned with the progression in time of state vectors in state space, describing a trajectory of the system's evolution with the potential for bifurcation. Citing a few investigators who have adopted this route :Wackermann's (52) assessment of electroencephalographic field changes as State Space trajectories, Hobson's (53) view of different stages of wakefulness and sleep in terms of state space dynamics; the demonstration of global brain state transitions occurring simultaneously across multiple forebrain areas (32); and a mapping between brain states and phenomenal experience (54). Application of state space concepts in the form of Coordination Dynamics has significantly enriched the understanding of the relations between global brain dynamics and behavior (30).

The Introduction and Section 2 referred to changes of scale as source of a new physical reality, requiring new descriptors (55,56). How does this new physical reality come into being, and what does it entail ? At the critical point of state transition, the micro state undergoes a profound reconfiguration which, among other features, is expressed as change of the correlation among its elements. The correlation function characterizes how the value at one point in state space is correlated with the value at another point, reflecting the micro level's fine structure. While under stable conditions extending over short distances, correlation length progressively increases as the critical point of state transition is approached. At the critical point itself, correlation length diverges to the extent that only correlations extending over larger scales remain. This implies that the system, metaphorically speaking, looses a detailed 'memory' of its microscopic structure. Thus, the macroscopic manifestation is at the critical point essentially based on a kind of abstraction from the original micro level, with all but those micro level features preserved that now determine the novel macroscopic observables at a new scale. This is also the point of drastic reduction of microscopic degrees of freedom. The change in correlations among the microscopic features at state transition can also be viewed as change to a coarser state space topology with new neighborhood relations among features, and thus associated with novel physical manifestations. Concurrently, the microscopic structure looses any characteristic length scale for system specific variables: it becomes scale invariant, i.e. fractal . Recent observations of Chialvo et al (57) point to the relevance of these considerations for brain processes: applying the technique of Fox et al (58) voxel based correlations of BOLD activity of different brain regions were obtained in fMRI studies of humans (59). These correlation maps were similar to those obtained computationally with Ising models in critical state transitions displaying long-range spin correlations. This observations support the notion that long range correlations among neural groups do obtain in the brain, as sign of it being in critical state.

Thinking in terms of State Space directs attention to the elaborate theoretical framework of state transitions. For a principled approach, it is here proposed that the Physics of Critical State Transitions (55,56) offers useful guidelines for relating macroscopic to microscopic system properties. In this framework, the previously noted abruptness of large-scale changes of the neural state space configurations would reflect a state transition to a new level of organization. Note the fundamental tenet of Critical Theory, in distinction from flow of trajectories in state space: Critical Theory is concerned with the origin of genuinely novel organizational patterns, triggered when system parameters attain critical values. This framework constitutes an alternative to conventional views of "information flow" along neural pathways and relay stations (60). Instead, transitions between levels of organization would be punctuated by abrupt transitions, such that consecutive levels of organization emerge as coarse grained abstractions of another level; each level being a distinct collective state of matter expressed in terms of its own organization on its own intrinsic scale. Rules governing the behavior of one level are no longer valid at the next level.

Generalizing beyond Neuroscience: in the view of Critical Theory, reality is composed of a hierarchy of levels and scales: at the state transitions, an intrinsically new scale emerges at each level, differing from that of the next finer level by ignoring some of the (irrelevant) details of the latter (13, 14). In this approach, the so called higher level description is not an evolution, nor an approximation of the fundamental (lower) level state, but represents a qualitatively new pattern of reality (61).

The emphasis in this essay on illustrating the essential features of the conceptual approach of the Theory of Critical State Transitions to neural systems should not obscure the generality of its relevance to pattern formation in nonequilibrium dynamic systems: merely to sketch here the scope of applicability, I refer to issues in morphogenesis (62 , 63 ) transition dynamics of biological systems on mesoscopic scale (e.g.: 64 ), and case studies of human social systems (e.g.: 65 ). For a more comprehensive collection of relevant topics, see Ref. 66 .

**Some final Questions : what can Modern Critical Theory contribute to Neuroscience ?**

The foregoing discussion contends that the abrupt reconfiguration of neural state space may be the manifestations of global brain state transitions: that is, a qualitatively novel expression of functional organization at a new scale of reduced dimensionality. In this view, the process of state transition of the brain's complex dynamical system enlists the conceptual and interpretive repertoire of the theory of Critical State Transitions, in analogy to the Physics of Condensed Matter. Thinking in this framework raises numerous questions, the answers to which are beyond one's intuitive grasp, but are amenable to computational simulation: What are the "tipping points" (67) for neural state space transitions ? What is the space of potential reconfigurations that such systems can in principle undergo under perturbation ? Under what conditions do they sustain stability ?

Bear in mind that the main thrust of this essay is the notion of emergence of new patterns of reality, due to changes in scales. The evolution of complex dynamic systems that cannot be deduced from their microscopic configurations, but can, at best, be approximated by equivalence classes of microscopic models. It is then a pragmatic issue to select from among candidate models those with best predictive value for macrosystem performance, and in closest accord with features and constraints imposed by the system's known micro- and mesoscopic organization. This approach also raises a question about universality classes: Can it be shown that neural systems belong to a universality class, or are unique and in a class by themselves ? If so, on account of what property ? And conversely: do there exist micro level states of other matter which, on state transition, constitute a neural system like universality class ?

To come to appreciate the space of possibilities in brain dynamics, these kinds of questions warrant exploring in their own right with computational models applying methods of statistical mechanics . Points of departure are suggested by patterns of collective neuronal activity in cultured brain slices, attributed by Beggs and Plenz (68,69) and Haldeman and Beggs (70) to a critical branching process, and by Breskin et al. (71) to Percolation transitions. Beggs and Thiagarajan (72) discussed several theoretical models to account for the fractal structure of neuronal avalanches as manifestations of cell assemblies, with the characteristics property of being scale free. Additional reference points are: packages of neural spike ("synfire chains") exhibit critical behavior, corresponding percolation state transitions (73); ion channel dynamics has the signature of a percolating network (74). Kozma et al (75) applied Percolation models to explore the dynamics of neuropil. As a program of research, the framework adumbrated in this essay and by these observations would situate brains squarely into the domain of the Physics of Condensed Matter.

The State Space framework of the relation between finer- and coarser levels implies inevitably an intersection of Complexity Science with the perennial philosophical problem of Emergence (76). Kim (77) identified five main tenets of "the doctrine of Emergentism", singling out as defining features the coming-together of lower-level entities in new structural configurations; the origin of

"higher level" properties, their unpredictability and irreducibility and, finally, the causal efficacy of emergent properties of their own. In light of the foregoing discussion of the distinction between a finer and a coarser level, it appears that the decisive event at Critical State Transitions is the origin of a new scale and, consequently, the point at which new objects, properties and laws originate, commensurate to that new scale. Taking the view of reality as a hierarchy of scales as basis, objects and properties in reality appear at each level in the hierarchy with their own organizational laws and structure, and with the propensity to undergo sharp state transitions (13). States are cases of emergence, based on Nature having walls of scales. Changes in scales, in turn, define new Ontologies for new descriptions.

The recent comprehensive overview by Deco et al (78) affords the opportunity for bringing in sharper focus the differences in approach and outlook presented by these authors, and the view presented in this essay. In the former, two approaches are considered in the summary of numerical simulations: applying a multiscale hierarchy with self-consistent evolution equations at each scale for coupling the emergent dynamics from fine to coarser scales; or, alternatively, to recursively enslave micro- and mesoscale dynamics for generating macroscopic field oscillations driven by mean field synaptic currents. In the present speculative essay, the emphasis is radically different: here, the relation between levels of functional organization are punctuated and discontinuous, metaphorically comparable to a "bucket brigade", with each "bucket" processing its "content" to the point of a state transition ; thus, passing the result of its intrinsic processing at coarse-grained reduced dimensionality to a recipient level. Accordingly, what is being passed from level to level are the descriptions of objects generated at scales , intrinsic to each bucket. Whether the latter view, with its conceptual inheritance from Condensed Matter Physics, is by itself or possibly in combination with other dynamic approaches of any merit, is a matter of future inquiry.

**Summary**.

The conceptual principles of Critical State Transitions are discussed, with the relation between neural events thought to constitute Cognition and Consciousness serving as an illustrative case study. The focus is on the origin of new scales in state transitions, and their role for defining emergent levels of Ontologies and descriptions with reduced dimensionality. The intent is to give credence to the notion that the approach here illustrated for the neural events in Cognition and Consciousness can provide valuable insights for relations among all levels of the nervous system's functional architecture.